\documentclass[
preprint,
superscriptaddress,
 amsmath,amssymb,
 aps,
floatfix,
showkeys,
]{revtex4-2}

\usepackage{graphicx}
\graphicspath{{figures/}} 
\usepackage{xcolor}
\usepackage{dcolumn}
\usepackage{bm}
\usepackage{hyperref}
\usepackage{longtable}
\usepackage{enumitem}
\usepackage{cleveref}
\usepackage[caption=false]{subfig}
\usepackage{verbatim}
\usepackage{multirow}
\usepackage{amssymb}

\newcommand{\gn}{\ensuremath{G^\textrm{3}_i}}
\newcommand{\gw}{\ensuremath{G^\textrm{9}_i}}
\begin{document}

\title{Machine learning potential for the Cu-W system}

\author{Manura Liyanage}
\email{manura.liyanage@empa.ch}
\author{Vladyslav Turlo}
\affiliation{Laboratory for Advanced Materials Processing, Empa - Swiss Federal Laboratories for Materials Science and Technology, Thun, Switzerland}%
\author{W. A. Curtin}%
\affiliation{National Centre for Computational Design and Discovery of Novel Materials MARVEL, \'{E}cole Polytechnique F\'ed\'erale de Lausanne, CH-1015 Lausanne, Switzerland}%
\affiliation{School of Engineering, Brown University, Providence, RI 02906 USA}%
\date{\today}





\begin{abstract}
Combining the excellent thermal and electrical properties of Cu with the high abrasion resistance and thermal stability of W, Cu-W nanoparticle-reinforced metal matrix composites and nano-multilayers (NMLs) are finding applications as brazing fillers and shielding material for plasma and radiation.  Due to the large lattice mismatch between fcc Cu and bcc W, these systems have complex interfaces that are beyond the scales suitable for \textit{ab initio} methods, thus motivating the development of chemically accurate interatomic potentials. Here, a neural network potential (NNP) for Cu-W is developed within the Behler-Parrinello framework using a curated training dataset that captures metallurgically-relevant local atomic environments. The Cu-W NNP accurately predicts (i) the metallurgical properties (elasticity, stacking faults, dislocations, thermodynamic behavior) in elemental Cu and W, (ii) energies and structures of Cu-W intermetallics and solid solutions, and (iii) a range of fcc Cu/bcc W  interfaces, and exhibits physically-reasonable behavior for solid W/liquid Cu systems.  As will be demonstrated in forthcoming work, this near-\textit{ab initio}-accurate NNP can be applied to understand complex phenomena involving interface-driven processes and properties in Cu-W composites. 
\end{abstract}


\maketitle

\section{\label{sec:introduction}Introduction}

The alloying of Cu, having high thermal and electrical conductivity, and W, having low thermal expansion coefficient, high elastic moduli, high hardness, good thermal stability, and abrasion resistance \cite{Kong2002,Tsakiris2014,Thomas2017,Wei2019,Li2023} leads to Cu-W composite materials with outstanding properties \cite{Petrovskiy2020,Lorenzin2022}. A highly positive heat of formation between Cu and W makes them immiscible under equilibrium conditions \cite{Kong2002,Gong2003,Zhang2003,Zhang2004,Wei2019,Zeng2021}, although direct alloying of Cu-W has been realized through non-equilibrium processing techniques such as mechanical alloying, ion beam mixing, and direct diffusion methods, all used to improve the bonding strength of the Cu-W interfaces \cite{Thomas2017,Zeng2021}. Cu-W composites with $\mathrm{\mu m}$-sized grains of Cu and W can be used in electric resistance welding, electrical contacts, and electro-discharge machining electrodes \cite{Li2023,Ubale2018}, heat sinks in fusion reactors \cite{Tejado2018,Tejado2020}, as well as power packaging in microelectronics and optoelectronics \cite{Ho2008}.  

Nano-multilayers (NMLs) are a special kind of composite that can be manufactured with physical vapor deposition methods and are characterized by a large density of interfaces. Experimental and computational work on fcc-bcc multilayers has shown the potential of NML by acting as obstacles to slip and sinks for radiation-induced defects \cite{Misra2001,Misra2005,Misra2007}.  Cu-W NMLs specifically, have strong potential to be used as brazing fillers and shielding material for radiation and plasma \cite{Girault2006,Cancellieri2016,Moszner2016,Moszner2016a,Druzhinin2019,Druzhinin2021,Druzhinin2021a,Lorenzin2022}. For brazing applications, Cu-W NMLs take advantage of microstructural evolution to a nanocomposite structure of W particles in a Cu matrix occurring at $\sim$700-800 $^\circ$C \cite{Cancellieri2016,Moszner2016,Moszner2016a,Druzhinin2019,Druzhinin2021,Druzhinin2021a,Li2022,Troncoso2024} - well below the melting points of Cu and W, while for shielding applications the NML interfaces act as sinks for radiation-generated defects \cite{Liang2013,Monclus2014,Callisti2016,Ma2018}.

Experimental observations have confirmed a strong bias toward forming interfaces with $\{111\}_\mathrm{Cu}||\{110\}_\mathrm{W}$ orientation relationship as expected for NMLs made of fcc (face-centered cubic) and bcc (body-centered cubic) metals with a strong fiber texture \cite{Taylor1966,Moss1969,Gadyszewski1994,Cancellieri2016,Moszner2016,Druzhinin2019,Yang2021,Yeom2023}. These interfaces are expected to be highly incoherent due to the large lattice mismatch between Cu and W. However, there are conflicting views regarding the in-plane orientation relationship with the Nishiyama-Wasserman (NW) \cite{Taylor1966,Reshoft1999} and Kurdjumov-Sachs (KS) \cite{Monclus2014, Druzhinin2019,Druzhinin2021a} misorientations, common for bcc/fcc NMLs, reported as possible candidates. \citet{Ma2018} state that the KS orientation relationship possesses the lowest possible interface energy, which contradicts \textit{ab initio} simulations in \citet{Bodlos2022} and a previous machine learning potential (MLP) \cite{Lorenzin2024}. Intermixing and alloying at the interface \cite{Melmed1965,Taylor1966,Wen2007,Zeng2021} and the adsorption of bcc Cu on W surfaces (up to 3 layers) \cite{Bauer1974} have also been observed in experiments; these highlight the complex nature of Cu-W interfaces, with structural and chemical features strongly affecting Cu-W NML properties. 

The level of detail provided by experimental work at the micro and nanoscales is limited in revealing the atomistic mechanisms occurring at Cu-W interfaces. Although \textit{ab inito} methods such as Density Functional Theory (DFT) have been used to study and predict some behavior of Cu-W composites \cite{Bodlos2022,Liang2013,Troncoso2024,Lorenzin2024} and metastable intermetallics \cite{Zhang2004,Liang2017}, results are limited to small computationally-feasible cell sizes that cannot capture more-realistic large incoherent interfaces and nanometric layers. Semi-empirical potentials such as the embedded atom method \cite{Daw1983,Daw1984} and the extended Finnis-Sinclair formalism\cite{Finnis1984,Tai2006} have thus been developed to model melting and diffusion at Cu-W interfaces \cite{Zeng2021}, Cu-W solid solutions \cite{Gong2003,Zhang2004,Yang2022}, glass forming capability \cite{Gong2003}, and interfacial reactions \cite{Kong2002,Wei2019}.  However, the small number of fitting parameters based on a limited DFT database composed of metastable intermetallic compounds prevents such models from capturing a more comprehensive range of chemical and structural features, such as interfacial defects, alloying at the interfaces, and interactions between the interface and defects in Cu or W.

Machine learning methods are now allowing the creation of machine-learned interatomic potentials (MLPs) that provide near-DFT accuracy at computational costs approaching traditional semi-empirical methods.  Following successes in other metals and alloys (Al-Cu \cite{Marchand2020}; Mg \cite{Stricker2020}; Al-Mg-Si \cite{Jain2021}; Al-Mg-Cu-Zn \cite{Marchand2022}; Zr \cite{Liyanage2022}), here we develop and validate an MLP using the neural network formulation of Behler and Parinello \cite{Behler2007,Behler2011}. This Cu-W neural network potential (NNP) captures complex Cu-W interfaces, solid solutions, intermetallics, and important phenomena in fcc Cu and bcc W, including lattice parameters, elastic parameters, stacking faults, surfaces, and dislocations.  We then show a few preliminary large-scale applications to Cu-W situations relevant to NMLs and NCs.  

While beyond the scope of this paper, forthcoming work will apply the NNP to study a wide range of application-oriented issues in NMLs.  Atomistic simulations using the NNP will be shown to help understand intermixing, phase stability, and observed large variations in interface stresses versus processing conditions \cite{Hu2024}.  Similarly, the NNP will be used to clarify the effects of layer thickness, in-plane stress, macroscopic porosity, temperature, grain boundaries, and non-equilibrium vacancies in physical vapor-deposited Cu/W nano-multilayers on Young's modulus to explaining contradictory experimental measurements versus deposition conditions and resulting microstructures \cite{Lorenzin-elastic2024}.  More broadly, similar NNPs could be developed for other fcc/bcc systems that are of high interest as NML structures \cite{Zhang1999,Choe2000,Misra2001,Misra2005,Misra2007,Srinivasan2007}.

The remainder of this paper is organized as follows.  \Cref{sec:methods} presents the formation of the reference dataset, a summary of the Behler-Parrinello framework, and the selection of the descriptors for the NNP. The quality of NNP for properties of elementary Cu and W is assessed in \cref{sec:unitary}.  In \cref{sec:binary}, the behavior of the NNP for binary Cu-W systems, including substitutions, metastable phases, Cu-W interfaces, and solid Cu/liquid W interactions, is investigated. Conclusions are presented in \cref{sec:conclusions} along with future applications of the new NNP.

\section{Methods\label{sec:methods}}

The following sequence of steps needs to be followed with care to develop a precise and useful NNP:
\begin{itemize}
    \item Development of a comprehensive reference database of structures, using \textit{ab intio} methods. This database should contain energy and force descriptions for unitary and binary systems and be complete in terms of local atomic environments encountered in possible applications;
    \item Selection of a set of descriptors. These descriptors play a crucial role in discerning the local atomic environments in the reference dataset. They should be sufficient in describing these environments without causing overfitting, a key consideration in the NNP development process;
    \item Training of the NNP on the reference dataset using the selected descriptors by optimizing the weights and biases in the neural network;
    \item Meticulous and comprehensive assessment of the quality of the final NNP. This includes thorough checks for overfitting and complex material behavior, such as dislocations, phase stability, and interfaces. 
\end{itemize}

Special care has to be taken to identify and avoid unphysical behavior while complying with underlying theories and models.  We will discuss how the above steps were executed in this work to create the Cu-W NNP.  Since the literature has comprehensively addressed NNP generation in metal alloys \cite{Marchand2020,Stricker2020,Jain2021,Marchand2022,Liyanage2022}, we discuss the framework only briefly.

\subsection{Reference structures}

The reference first-principles DFT dataset must be constructed carefully to include all local atomic environments expected to be encountered in future applications of the NNP. \citet{Marchand2020} has created an extensive dataset for fcc Cu behavior \citet{Marchand2020a} that captures lattice parameters, equation of states ($\pm20\%$ lattice parameter), elastic behavior (for strains $\pm0.2\%$), stacking faults, vacancies, interstitials, and vicinal surfaces. We have created a similar dataset for W, capturing the linear and non-linear behavior of the bcc W. For Cu-W solid solutions, metastable alloys with space groups B$_\mathrm{h}$, B$_3$, L1$_0$, and L1$_1$ for the CuW stoichiometry, space group C15 for Cu$_2$W and CuW$_2$ stoichiometries, and space groups A15, C15, D0$_3$, D0$_9$, D0$_{19}$, D0$_{22}$, and L1$_2$ for Cu$_3$W and CuW$_3$ stoichiometries, along with the intermetallic structures available in the Open Quantum Material Database (OQMD) database \cite{Kirklin2015} are included in the equilibrium configurations. Structures for the space groups A15, B2, D0$_9$, D0$_{19}$, D0$_{22}$, and L1$_2$, identified as more likely, are studied under strains of 0.15\% for the Voigt directions.  Cu and W are substituted in bcc W and fcc Cu, respectively, as single atoms and pairs of atoms at 1st, 2nd, and 3rd nearest neighbor distances.  Cu-W interfaces are modeled in compact cells achievable with DFT for $\{100\}_\mathrm{Cu}||\{100\}_\mathrm{W}$ (two orientations with 45 inclination),  $\{110\}_\mathrm{Cu}||\{110\}_\mathrm{W}$, $\{111\}_\mathrm{Cu}||\{110\}_\mathrm{W}$ (two orthogonal orientations), and $\{111\}_\mathrm{Cu}||\{111\}_\mathrm{W}$ interfaces. We examine the equilibrium configurations and structures having relative slip across the Cu-W interfaces.  We also generate structures for the experimentally observed $\{111\}_\mathrm{Cu}||\{110\}_\mathrm{W}$ orientations by swapping Cu and W atoms in the 1st, 2nd, and 3rd layers away from the interfaces.  We include structures relevant to thermal vibrations by studying fcc Cu and bcc W with random atom displacements away from the equilibrium positions. Random displacements have a normal distribution with a mean of 0 and a standard deviation of 0.15 \r{A}.

The reference dataset contains 4583 structures with 136,928 atomic environments (78,436 around Cu atoms and 58,492 around W atoms) and 410,784 force components along the cartesian directions. The additional dataset created in the current study is openly available on the Materials Cloud \cite{dataset}. 

\subsection{DFT methodology}

First-principles calculations are performed with the DFT framework of Quantum Espresso  \cite{Giannozzi2009}. We adopt the DFT parameters in \citet{Marchand2020} to be consistent with the available Cu reference dataset. The standard generalized gradient approximation of \citet{Perdew1996} (GGA-PBE) described the electron-electron exchange-correlation effects. Basis sets for Cu and W are $3p$ $3d$ $4s$ and $5s$ $5p$ $5d$ $6s$ states, respectively. Plane-wave cutoff energy of 544 eV (40 Ry) and a $\Gamma$-centered Monkhorst-Pack \textbf{k}-point mesh \cite{Monkhorst1976} with a spacing of 1/80 r{A}$^{-1}$ are used with a Gaussian integration scheme with a broadening of 0.6 eV.  The Broyden–Fletcher–Goldfarb–Shanno algorithm optimization was used to ionic and simulation cell relaxations \cite{Broyden1970,Fletcher1970,Goldfarb1970,Shanno1970}.

\subsection{Neural-network framework and symmetry functions}

Various neural network frameworks for the creation of reference structures (such as active learning), descriptors, and training algorithms are extensively discussed in the literature \cite{Mishin2021}.  Here, we use the Behler-Parinello descriptors that represent the local atomic environments through atomic pairs and triplets with radial and angular symmetry functions.  The open-source code \texttt{n2p2} \cite{Singraber2019} is used to train the NNP and is then also implemented in LAMMPS \cite{Plimpton1995} in further studies and applications. We highlight the main aspects here.

NNPs consider the energy of an atomic ensemble to be the aggregate of individual atomic energies. Atomic energies depend primarily on their local atomic environments, described by the atom-centered symmetry functions selected based on the atomic coordinates of the reference dataset. Three types of Behler-Parrinello SFs (radial ($G^2_i$) SFs for pairs; narrow angular (\gn); wide angular (\gw) SFs for triplets) are utilized when developing the NNP. For a central atom $i$, these SFs have the forms
\begin{equation}\label{eq:radial}
    G_i^2 = \sum_{j\neq i} e^{-\eta(r_{ij}-r_s)^2}f_c\left( r_{ij} \right),
\end{equation}
\begin{equation}\label{eq:narrow}
\begin{aligned}
    G_i^3= &2^{1-\zeta}\sum_{j,k\neq i} (1+\lambda \cos{\theta_{ijk}})^\zeta e^{-\eta\left(r_{ij}^2+r_{ik}^2+r_{jk}^2\right)}\\ 
    &\times f_c\left(r_{ij}\right)f_c\left(r_{ik}\right) f_c\left(r_{jk}\right),
\end{aligned}
\end{equation}
\begin{equation}\label{eq:wide}   
\begin{aligned}
    G_i^9 = &2^{1-\zeta}\sum_{j,k\neq i} (1+\lambda \cos{\theta_{ijk}})^\zeta e^{-\eta\left(r_{ij}^2+r_{ik}^2\right)} \\ &\times f_c\left(r_{ij}\right) f_c\left(r_{ik}\right),
\end{aligned} 
\end{equation}
 where $r_{ij}=|\mathbf{r_j}-\mathbf{r_i}|$ is the radial distance between atom $i$ and atom $j$ in the local environment and $\theta_{ijk}$ is the angle between the vectors from atom $i$ to atoms $j$ and $k$, respectively, capturing the geometries of the neighboring atoms in the local environments.  The hyperparameters $\eta$, $r_s$, $\zeta$, and $\lambda$ are selected through sampling methods proposed in \citet{Gastegger2018} and \citet{Imbalzano2018}.  $f_c$ is a cutoff function that reduces the radial components to zero smoothly at a radial cutoff $r_\textrm{c}$, and here we use  $f_c(r)=\tanh^3\left(1-r/r_\textrm{c}\right)$. The most useful SFs describing the structures in the reference dataset were then obtained with the PCov-CUR (principle covariant regression CUR) algorithms \cite{Imbalzano2018,Cersonsk2021,Goscinski2023}.

With the descriptors (symmetry functions) identified, the energy $E_i$ of atom $i$ in its local environment is calculated as
\begin{equation}\label{eq:energy_NN}
\begin{aligned}
    E_i &= f_1^3\left\{b_1^3+\sum_{k=1}^{M_{\mathrm{layer},2}}w_{n1}^{23} f_n^2 \right.\\
    &
\times\left.\left[b_n^2+\sum_{m=1}^{M_{\mathrm{layer},1}}w_{mn}^{12}f_m^1\left(b_m^1+\sum_{l=1}^{M_{\mathrm{sym}}}w_{lm}^{01}G_{i,l}\right)\right]\right\},
    \end{aligned}
\end{equation}
where $G_{i,l}$ is the value of the symmetry function around atom $i$ for the $l^{th}$ set of hyperparameters, $f(\cdot)$ are the activation functions for each node (for the current study $f_m$ and $f_n$ are softplus functions and $f_1$ is the identity function), $M_{\mathrm{layer},p}$ is the number of nodes in the $p^{th}$ hidden layer, $M_{\mathrm{sym}}$ is the number of SFs. The weights $w^{uv}_{pq}$ connect the $p^{th}$ node in the $u^{th}$ layer to the $q^{th}$ node in the $v^{th}$ layer and biases $b^v_q$ are added to the $q^{th}$ node in the $v^{th}$ layer of the neural network.  Training is carried out to optimize the weights and biases of the neural network through an iterative process to obtain the best fit to obtain the energy and forces for the structures in the training dataset. The optimization minimizes the loss function
\begin{equation}\label{eq:loss_func}
\Gamma = \sum^{N_\mathrm{struct}}_{n=1}(E_{\mathrm{NN}}^n-E_{\mathrm{ref}}^n)^2+\beta^2\sum^{N_\mathrm{struct}}_{n=1}\sum_{m=1}^{3N_{\mathrm{atom}}^n}(F_{n,\mathrm{NN}}^m-F_{n,\mathrm{ref}}^m)^2,
\end{equation}
where $E$ and $F$ are the energies of the structures and the forces of the atoms in each structure, respectively, with the subscripts ``ref" and `NN" for DFT reference values and the NNP predictions. The parameter $\beta$ (with \r{A} units) is used to scale the weight of the forces (which are in units of eV/\r{A}) relative to the energies (in eV).

Infinite combinations of symmetry functions are available to describe atoms' local atomic neighborhoods. However, the selection will decide the accuracy, efficiency, and possibility of overfitting. Hence, the symmetry functions need to be chosen carefully. Symmetry functions in \citet{Marchand2020} for Cu did not provide very good predictions regarding the interfaces. Hence, we used $r_c=7.0$ \r{A}, $\lambda=\left[1,-1\right]$, and $\zeta=\left[1,3,12,64\right]$ with a sampling of $\eta$ and $r_s$ as defined by \citet{Gastegger2018} and \citet{Imbalzano2018}.  This led to 326 symmetry functions each for Cu and W, respectively. Using the PCov-CUR decomposition algorithms and geometries of the input structures, we selected the 48 "most useful" SFs (independent SFs that least affect feature identification) for Cu and for W to describe the local atomic environments in the training set. 

The training uses a Kalman filter optimization carried out for a fixed number of 400 epochs. The training dataset is selected as 90\% of the reference dataset.  The remainder is kept to test the developed NNP for overfitting and any atomic environment of the reference dataset unrepresented in the training dataset.  2\% of the force components from the training dataset were used for training to prevent the force components (3 per atom) from overwhelming the energies (one energy per structure), with 100\% of the energies used to update the NNP per each epoch. For the current NNP, $\beta=1$ \r{A} is used for the force weighting in \cref{eq:loss_func} such that a 1 meV/\r{A} force error is given the same weight as a 1 meV energy error.

\section{Results and discussion\label{sec:results}}

An NNP must be examined for indications of overfitting, outliers in the testing dataset, and interplay of different atomic environments in the configurational space of the reference dataset. Further tests are then required to determine the suitability and accuracy of the NNP for any metallurgical applications. 20 NNPs were trained using different initial weights and biases, which didn't show any outliers, deficiency in the selection of the training dataset, or overfitting.  For each NNP, a root mean percentage square error (RMPSE) was calculated for the properties of fcc Cu, bcc W, and Cu-W interface structures. The NNP designated as "NNP01" in the supplementary material has the lowest RMPSE and shows the best overall behavior across the test configurations, and so it is the focus of our results for the remainder of this paper. The properties calculated for all the 20 NNPs are shown in the supplementary material.

\subsection{Structures in the reference dataset}

\begin{figure}[htpb]
    \subfloat[Errors in energy]{
    \includegraphics[width=0.5\textwidth]{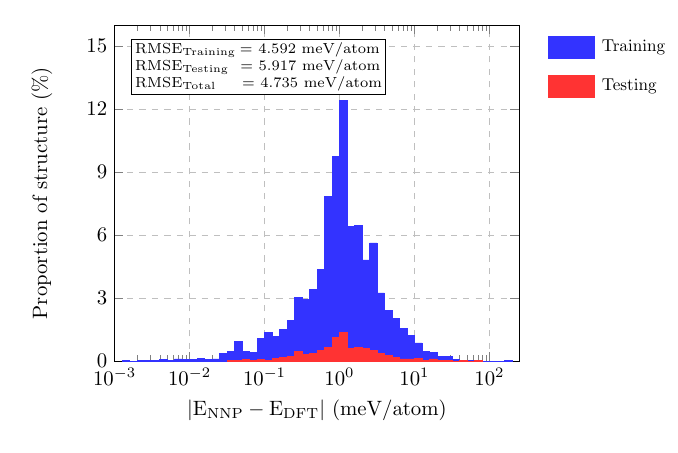}
    \label{fig:err_energy}}
    \subfloat[Error in forces]{
    \includegraphics[width=0.5\textwidth]{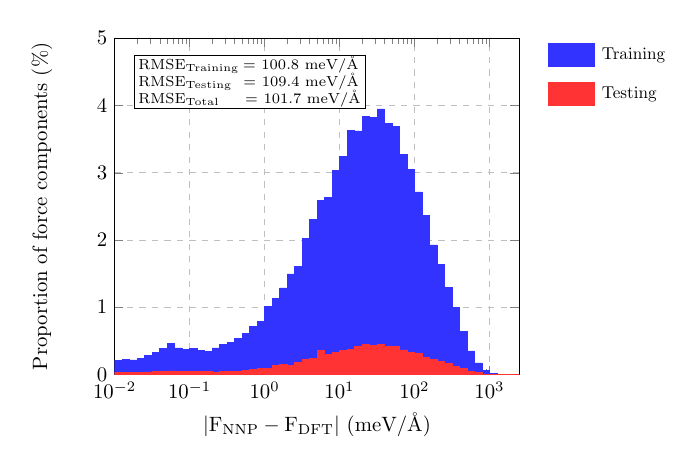}
    \label{fig:forces}}
     \caption{Histogram of errors for energies and forces of the structures in the training and testing datasets as predicted by the NNP. Values corresponding to the training and testing datasets are shown in blue and red,  respectively.}\label{fig:err_dist}
\end{figure}

The distribution of errors of energies and forces predicted by the NNP for the structures in the training test and testing set is shown in \cref{fig:err_dist}. The training and testing set errors are similar, with almost identical root mean square errors (RMSE). The RMSE for energy is higher than in other studies \cite{Marchand2020,Stricker2020,Jain2021,Marchand2022,Liyanage2022}, but still within a few meV/atom. Most structures with energy errors $>5$ meV/atom comprise metastable Cu-W and W with large strains ($V/V_0>20\%$). The RMSE would be comparable to other studies if these structures were neglected. Similar observations can be made for the force errors, with all force errors $>$400 meV/\r{A} corresponding to structures far from their equilibrium states, specifically unrelaxed structures with interstitials and structures with atoms displacements far from their relaxed configurations.  Similar observations are reported in other studies, with minimal effect on the linear and non-linear mechanical behavior \cite{Liyanage2022}.  

Further tests to examine the NNP performance versus DFT and analytical results are discussed in the following section.

\subsection{Elemental Cu and W\label{sec:unitary}}

\begin{table}[htpb]
\centering
\caption{\label{tab:properties_Element}Comparison of basic mechanical properties of Cu and W predicted by the NNP and by DFT.}
    \begin{tabular}{l c c| c c}
    \hline\hline
    \multirow{2}{*}{Property}&\multicolumn{2}{c}{Cu}&\multicolumn{2}{c}{W}\\ \cline{2-5}
    & DFT & NNP & DFT & NNP \\
    \hline
\multicolumn{5}{l}{\textbf{Lattice properties}}\\							
a (\r{A})	&3.625	&3.626	&3.185	&3.186	\\
V/atom (\r{A}$^3$)	&11.904	&11.914	&16.155	&16.164	\\
E$_{coh}$ (eV/atom)	&-3.781	&-3.779	&-10.627	&-10.628	\\
\multicolumn{5}{l}{\textbf{Elastic Properties (GPa)}}\\						
C$_{11}$	&180	&172	&535	&528	\\
C$_{12}$	&125	&127	&208	&199	\\
C$_{44}$	&80	&72	&150	&151	\\
\multicolumn{5}{l}{\textbf{Surface energy (J/m$^2$)}}\\					
\{100\}	&1.486	&1.481	&4.002	&3.999	\\
\{110\}	&1.571	&1.570	&3.244	&3.318	\\
\{111\}	&1.329	&1.323	&3.563	&3.641	\\							
\multicolumn{5}{l}{\textbf{Other properties}}\\							
E$_{vac}$ (eV)	&1.088	&1.020	&3.308	&3.512	\\
E$_{SIA}$ - octahedral (eV)	&3.332	&3.235	&12.153	&11.992	\\
E$_{SIA}$ - tetrahedral (eV)	&3.716	&3.525	&11.565	&11.053	\\
\hline
    \end{tabular}
\end{table}

\begin{figure}[htpb]
    \centering
    \subfloat[fcc Cu]{\includegraphics[width=0.4\textwidth]{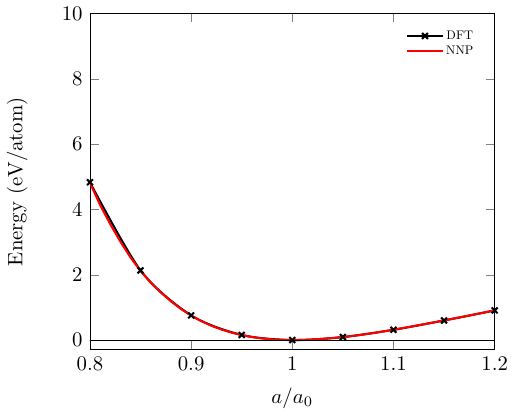}
         \label{fig:Cu_EOS}}
    \quad
    \subfloat[bcc W]{\includegraphics[width=0.4\textwidth]{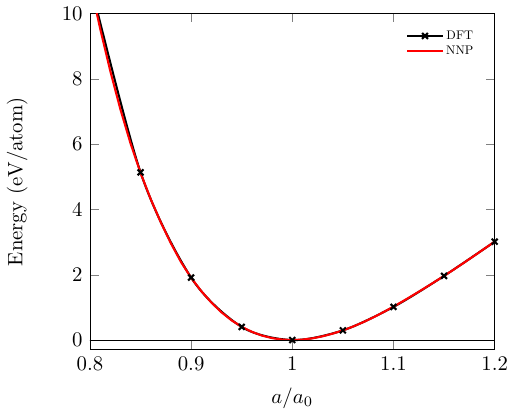}
         \label{fig:W_EOS}}
     \caption{Equation of states obtained for fcc Cu and bcc W with DFT and the NNP}\label{fig:CuW_EOS}
\end{figure}

The NNP shows excellent agreement with DFT for the properties shown in \cref{tab:properties_Element} with most errors $<10$\%. The lattice parameters, most of the elastic properties, and surface energies are in excellent agreement between DFT and The NNP. The shear modulus C$_{44}$ for Cu shows the most significant error of 11\%. However, this is a difficult property to predict accurately with NNPs, with other studies reporting more significant errors \cite{Marchand2020}. The vacancy formation energies (E$_\mathrm{vac}$) and self-interstitial energies (E$_\mathrm{SIA}$) show good agreement with the DFT values with maximum error $\sim$5\%. This indicates that the NNP models vacancies and interstitials with reasonable accuracy at the equilibrium states, even though the predicted forces of the non-equilibrium structures in the reference dataset have a noteworthy error. \Cref{fig:CuW_EOS} shows the equation of states for the fcc and bcc phases of Cu and W, respectively. The agreement is excellent for an extensive range of strains ($a/a_0=\pm$20\%).

\begin{figure}[b]
    \centering
    \subfloat[fcc Cu]{\includegraphics[width=0.45\textwidth]{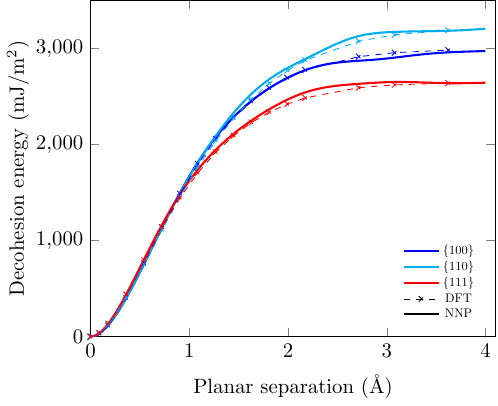}
         \label{fig:Cu_dec}}
    \qquad
    \subfloat[bcc W]{\includegraphics[width=0.45\textwidth]{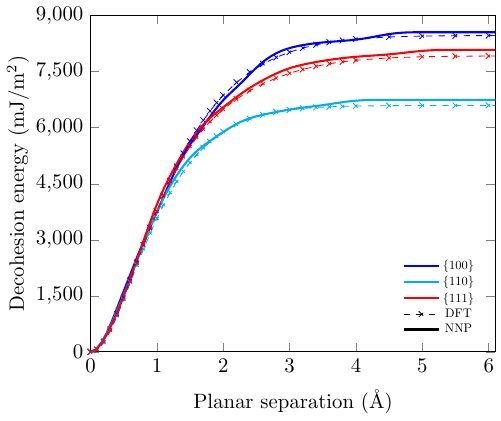}
         \label{fig:W_dec}}
    \caption{Decohesion curves for $\{100\}$,$\{110\}$, and $\{111\}$ families of planes for fcc Cu and bcc W comparing DFT and the NNP. Results from DFT are shown in dashed lines with $\times$, and results from the NNP are shown in solid lines. All DFT structures here are included in the reference dataset, with the non-DFT results corresponding to relaxed structures using the NNP}\label{fig:CuW_Dec}
\end{figure}

\begin{figure}[htpb]
    \centering
    \subfloat[(111) plane in fcc Cu with Burgers vector $\mathbf{b}=a/2\langle11\bar{2}\rangle$]{\includegraphics[width=0.45\textwidth]{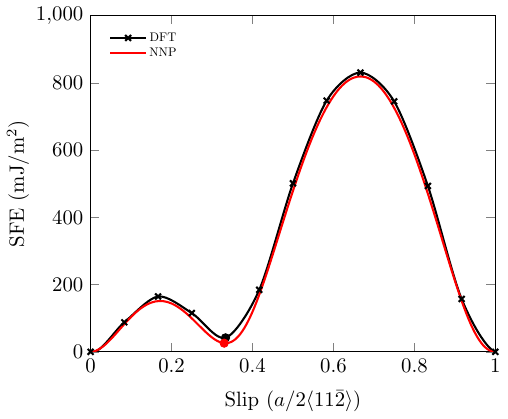}\label{fig:Cu_SF}}
    \qquad
    \subfloat[(110) plane in bcc W with Burgers vector $\mathbf{b}=a/2\langle11\bar{1}\rangle$]{\includegraphics[width=0.45\textwidth]{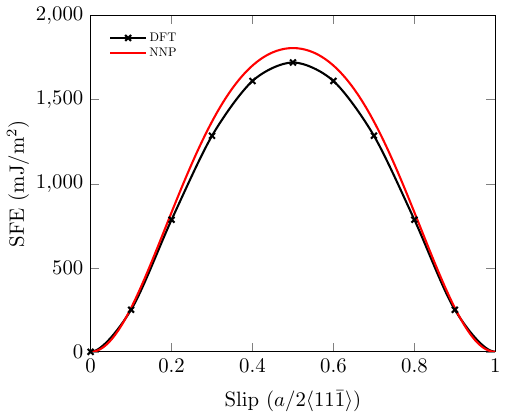}\label{fig:W_SF}}
    \caption{GSFE curves for the (111) plane in fcc Cu and the (110) plane in bcc W; comparing DFT and the NNP developed in the current study. * shows the location and energy of the stable stacking fault in fcc Cu. All DFT structures here are contained in the training dataset, with the non-DFT results corresponding to relaxed structures using the NNP}
\end{figure}

Decohesion and stacking fault energies are important in metallurgical applications to study plasticity and fracture. \Cref{fig:CuW_Dec} shows the decohesion curves for fcc Cu and bcc W for the $\{100\}$,$\{110\}$, and $\{111\}$ families of planes. The agreement between DFT and the NNP is excellent, especially for small displacements ($<$2 \r{A}). There are variances of $\sim2\%$ for large displacements, which have no significant effects in any application.  The Generalized Stacking Fault Energies (GSFE) for $\{100\}$, $\{110\}$, and $\{111\}$ planes for fcc Cu and bcc W are included in the training data, and overall agree well with DFT.  The most relevant cases for dislocations are the $\{111\}$ planes in Cu shown in \cref{fig:Cu_SF} and the $\{110\}$ planes in W shown in \cref{fig:W_SF}.  For Cu, the NNP unstable stacking fault energy is $\sim$1.5\% lower than the DFT value, which is sufficient accuracy for modeling dislocation emission from crack tips. However, the NNP stable stacking fault energy (SSFE) is $\sim40\%$ lower than the DFT value, which leads to a larger partial dislocation dissociation distance for the lattice dislocations (see below) and can strongly inhibit cross-slip.  Previous results for the Cu SSFE for NNPs developed using only the Cu dataset are 15-35\% lower than the DFT value \cite{Marchand2022}, so training here with the addition of W leads to slightly worse agreement.  For W, the unstable stacking fault energy estimated by NNP is just 5\% higher than the DFT value, which is sufficiently accurate to capture the emission behavior of crack tips. 

The accurate modeling of dislocation core structures is essential in understanding the plastic behavior of metals.  Dislocation core structures combine atomic environments related to elastic and stacking fault deformations.  The main operative slip planes are the most densely packed planes ($\{111\}$ for fcc Cu; $\{110\}$ for bcc W) and so we examine the edge and screw dislocations for these slip systems. Initial dislocation structures are created using the anisotropic Volterra elastic displacement field  \cite{Volterra1907} in cuboidal simulation cells with $30\times 30$ \r{A}$^2$ in-plane dimensions and one periodic unit along the dislocation line direction.  For dislocations in fcc Cu, several initial partial separation distances are used (up to 60 \r{A} in steps of 5 \r{A}). Atoms in a 10 \r{A} layer along the outer boundary are fixed and the remaining atoms relaxed to the minimum-energy configuration.  The relaxed dislocation core structures were analyzed with the Nye tensor \cite{Nye1953,Hartley2005}, differential displacement maps \cite{Vitek1970}, and disregistry along the slip plane using the \texttt{atomman} \texttt{python} package \cite{Atomman}.  Gradients of the disregistries enable understanding Burgers vector distribution along slip planes and provide an estimate of the partial separation distances \cite{Clouet2020}. A theoretical estimate for the partial separation distance can be obtained using the elastic properties and stable stacking fault energy along the slip direction as
\begin{equation}\label{eq:disl_dis}
    \mathrm{d}_\mathrm{theory} = \frac{\mathbf{b}^{(1)}_{i}\mathbf{K}_{ij}\mathbf{b}^{(2)}_{j}}{2\pi\gamma_{\mathrm{ssf}}},
\end{equation}
where $\mathbf{b}^{(1)}$ and $\mathbf{b}^{(1)}$ are the Burgers vectors of the partial dislocations, $\mathbf{K}$ is the Stroh matrix \cite{Stroh1958,Stroh1962}, and $\gamma_\mathrm{ssf}$ is the stable stacking fault energy of the respective plane. 

\begin{figure}[htpb]
    \centering
    \subfloat[Edge dislocation]{\includegraphics[width=0.45\textwidth]{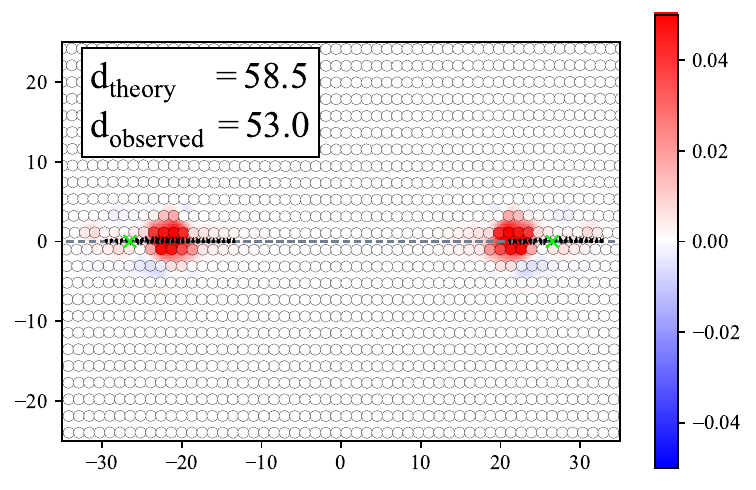}
         \label{fig:Cu_edge}}
    \qquad
        \subfloat[Screw dislocation]{\includegraphics[width=0.45\textwidth]{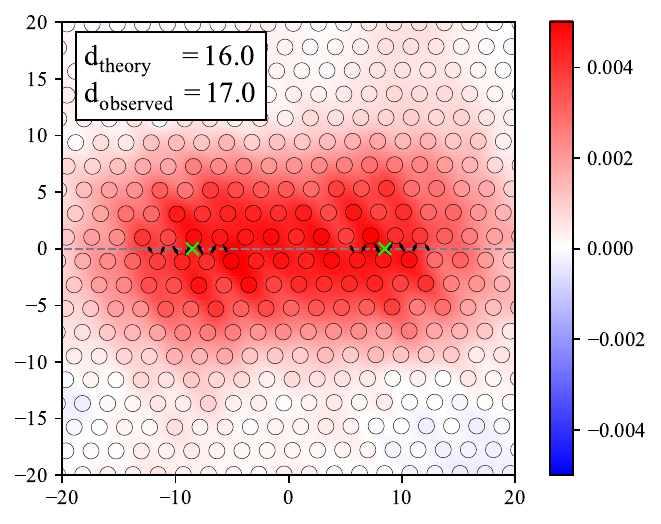}
         \label{fig:Cu_screw}}
     \caption{Atomistic core structures for $\mathbf{b}=0.5\langle 1\bar{1}0\rangle$ dislocations of $\{111\}$ planes in fcc Cu for the NNP, with the Nye tensors (for the dominant direction) and differential displacement maps overlaid. Green crosses indicate the centers of the partial dislocations identified by the disregistry plots. Theoretical and observed dislocation separation distances are indicated as $d_\mathrm{theory}$ and $d_\mathrm{observed}$ respectively.}\label{fig:Cu_dislocations}
\end{figure}

\begin{figure}[htpb]
    \centering
    \subfloat[Edge dislocation]{\includegraphics[width=0.45\textwidth]{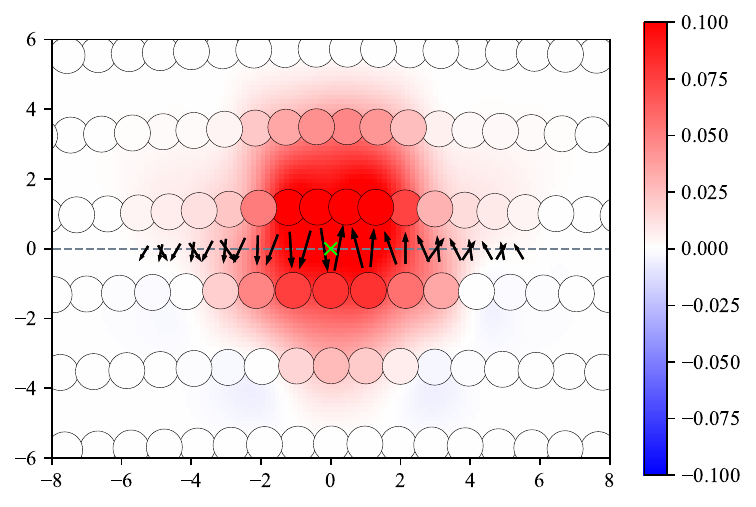}
         \label{fig:W_edge}}
    \qquad
    \subfloat[Screw dislocation]{\includegraphics[width=0.45\textwidth]{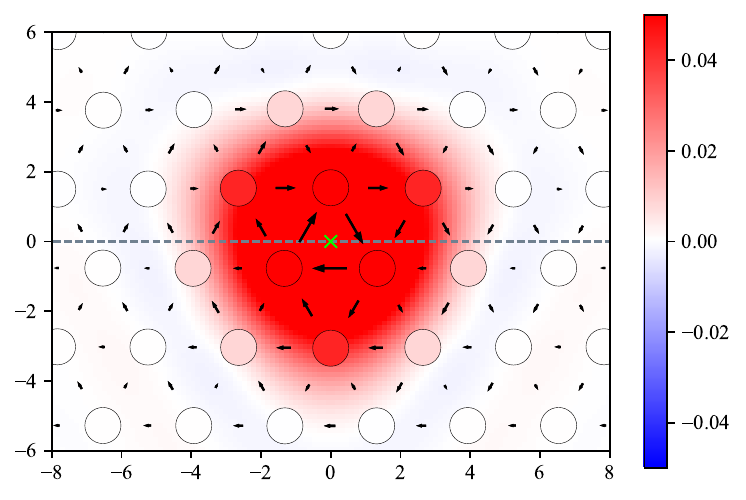}
         \label{fig:W_screw}}
     \caption{Atomistic core structures for $\mathbf{b}=0.5\langle \bar{1}11\rangle$ dislocations of $\{110\}$ planes in bcc W for the NNP, with the Nye tensor (for the dominant direction) and differential displacement maps overlaid. Green crosses indicate the centers of dislocations.}\label{fig:W_dislocations}
\end{figure}

The stable dislocation cores are shown in \cref{fig:Cu_dislocations,fig:W_dislocations} and exhibit physically-reasonable structures similar to those obtained previously using other potentials and/or DFT.  The dissociation distance in Cu agrees well with the theory, as indicated, computed using the NNP properties.  Using the DFT properties, d$_{theory}$ is $/sim$ 40 and 12 \r{A} for edge and screw dislocations, respectively, which is much smaller than the NNP-derived values due to the NNP error in the SSFE.  For W, the screw dislocation shows the expected compact symmetric structure and the edge dislocation shows a small spreading in the glide plane.  The NNPs are thus broadly suitable for modeling dislocation plasticity in Cu and W.

Due to the high temperatures involved in the manufacturing of Cu-W MMCs, accurate thermodynamic properties of Cu and W are essential for many applications.  Various properties are estimated using periodic arrangements of $8 \times 8 \times 8$ simulation cells (fcc and bcc unit cells for Cu and W, respectively) and executing molecular dynamics simulations in the isothermal–isobaric ensemble with a time step of 0.5 fs.  Before evaluating any properties, a system is equilibrated for 50,000 time steps (25 ps).  Melting points are estimated using the two-phase method, where the migration of the solid/liquid interface is monitored, with no net motion at the melting temperature. The heat of fusion E$_\mathrm{HF}$ is the energy absorbed to change the metal from solid to liquid at the melting temperature. The thermal expansion coefficient of the crystalline phase at T=300K was determined using the lattice constants measured in simulations at 250 K, 300 K, and 350 K.  The above simulated properties are compared with experimental observations in \cref{tab:thermo_Element}, showing good agreement.  The melting points of Cu and W are estimated as $1245\pm5$ K and $3537\pm13$ K, respectively, which are 100-150 K lower than experimental values.  The NNP estimates of E$_\mathrm{HF}$ show errors of $\sim 20\%$ and the thermal expansion values differ by roughly similar factors. Overall, these estimates are satisfactory since no liquid or high-temperature configurations are included in the reference training dataset.

\begin{table}[htpb]
\centering
\caption{\label{tab:thermo_Element}Comparison of the thermodynamic properties obtained for the trained NNP potential for Cu and W with the corresponding experimental properties.}
    \begin{tabular}{l c c| c c}
    \hline\hline
    \multirow{2}{*}{Property}&\multicolumn{2}{c}{Cu}&\multicolumn{2}{c}{W}\\ \cline{2-5}
    & Exp. & NNP & Exp. & NNP \\
    \hline
T$_\mathrm{m}$ (K) & 1358 & 1245 & 3687 & 3538 \\
E$_\mathrm{HF}$ (kJ mol$^{-1}$) &13.26 \footnotemark[1] & 11.51 &52.31 \footnotemark[1]\footnotemark[2] & 63.93\\
$\alpha_{300K}$ ($\times 10^{-6}$) &16.76 \footnotemark[3] & 20.4 &4.6 \footnotemark[4] & 4.3\\
\hline
    \end{tabular}
\footnotetext[1]{\citet{Dean1999}}
\footnotetext[2]{\citet{Lide2003}}
\footnotetext[3]{\citet{Wang1996}}
\footnotetext[4]{\citet{Hidnert1925}}
\end{table}

\subsection{Cu-W systems\label{sec:binary}}

Cu and W mainly form metal composites due to the high positive mixing enthalpy under equilibrium conditions, possibly with complex interfaces  \cite{Melmed1965,Bauer1974,Taylor1966,Wen2007,Zeng2021}.  However, Cu-W can form metastable intermetallic structures under non-equilibrium conditions \cite{Kong2002,Gong2003,Zhang2003,Zhang2004,Wei2019,Zeng2021}.  Some manufacturing processes involve liquid Cu and crystalline W. To assess the use of the NNP for applications across relevant situations arising in the processing and applicatin of Cu-W systems, we examine NNP predictions for (i) substitution of Cu in W and W in Cu, (ii) the properties of many metastable alloys, (iii) solid-state interfaces between fcc Cu and bcc W, and (iv) interactions between solid W and liquid Cu. 

\subsubsection{Solute atoms}

Although Cu and W are largely immiscible, some mixing has been observed in Cu-W NMLs at high temperatures.  The solute solution energy energy E$\mathrm{^{1sol}_{X}}$ (X=Cu or W) is defined as
\begin{equation}\label{eq:sub}
    \mathrm{E^{1sol}_{X}} = \mathrm{E^{1sol}_{(N-1)Y,X}}-\mathrm{E_{X}}-\mathrm{(N-1)E_{Y}}
\end{equation}
where E$_\mathrm{^{1sol}_{(N-1)Y,X}}$ is the total energy of the equilibrium simulation cell containing a single solute and N-1 atoms of the solvent Y=W or Cu. E$_\mathrm{X}$ and E$_\mathrm{Y}$ are the energies of solute and solvent atoms in their equilibrium phases (fcc Cu or bcc W). Solute-solute interaction, or binding, energies E$^\mathrm{bind}_\mathrm{X}$ are also examined for solute pairs (of X=Cu or W) at first, second, and third nearest-neighbor locations (1 NN, 2 NN, and 3 NN, respectively), defined as 
\begin{equation}\label{eq:binding}
    \mathrm{E^{bind}_{X}}=\mathrm{E^{2sol}_{(N-2)Y,2X}}-2\mathrm{E^{1sol}_{(N-1)Y,X}}+\mathrm{N\ E_{Y}}
\end{equation}
with the same definitions as \cref{eq:sub}. The necessary energies are computed using fully-periodic $5\times5\times5$ simulation cells (fcc and bcc unit cells for Cu and W, respectively). A negative E$^\mathrm{bind}_\mathrm{X}$ indicates attractions. Results for E$\mathrm{^{1sol}_{X}}$ and E$^\mathrm{bind}_\mathrm{X}$ obtained with the NNP and DFT are shown in \cref{tab:binding}.
The agreement between DFT and the NNP is good,  with the high positive values of E$\mathrm{^{1sol}_{X}}$ consistent with the high immiscibility.  Correspondingly, the solute-solute interactions are strongly attractive at first neighbors, consistent with the high immiscibility that drives phase separation into Cu and W.  The NNP binding energies at 2 NN and 3 NN show some differences with DFT (including cases with a change in sign), but overall, the values are very small compared to the 1 NN attraction and so have no significant consequences on physical behavior.

\begin{table}[htpb]
\centering
\caption{\label{tab:binding}E$\mathrm{^{1sol}_{X}}$ and E$^\mathrm{bind}_\mathrm{X}$ obtained with the NNP for substitution of Cu in bcc W and W in fcc Cu in eV}
    \begin{tabular}{l c c| c c}
    \hline\hline
    \multirow{2}{*}{Energy}&\multicolumn{2}{c}{Cu in bcc W}&\multicolumn{2}{c}{W in fcc Cu}\\ \cline{2-5}
    & DFT & NNP & DFT & NNP \\
    \hline
E$\mathrm{^{1sol}_{X}}$ & 1.464 & 1.568 & 1.876 & 1.816 \\
\textbf{E$^\mathrm{bind}_\mathrm{X}$}&&&&\\							
1 NN &-0.336&-0.479&-0.823&-0.728\\
2 NN &0.070&0.058&0.016&-0.041\\
3 NN &0.070&-0.081&-0.082&-0.017\\
\hline
    \end{tabular}
\end{table}

\subsubsection{Metastable intermetallic compounds and solid solutions}

\begin{figure*}[htpb]
    \centering
    \subfloat[A15 Cu$_3$W]{\includegraphics[width=0.2663\textwidth]{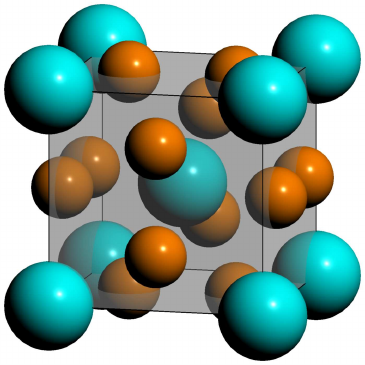}
         \label{fig:A15_Cu3W}}
    \qquad\qquad
    \subfloat[B$_2$ CuW]{\includegraphics[width=0.1696\textwidth]{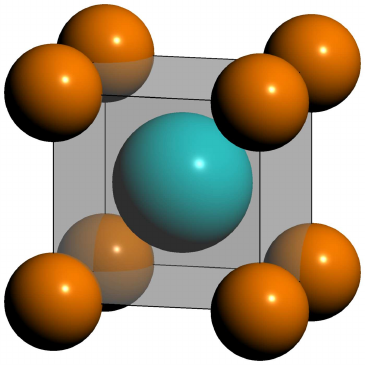}
         \label{fig:B_2_CuW}}
    \qquad\qquad
    \subfloat[D0$_{19}$ Cu$_3$W]{\includegraphics[width=0.3000\textwidth]{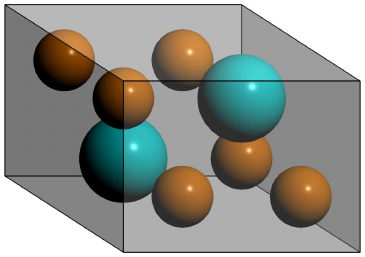}
         \label{fig:D0_19_Cu3W}}
    \qquad
    \subfloat[D0$_{22}$ Cu$_3$W]{\includegraphics[width=0.2211\textwidth]{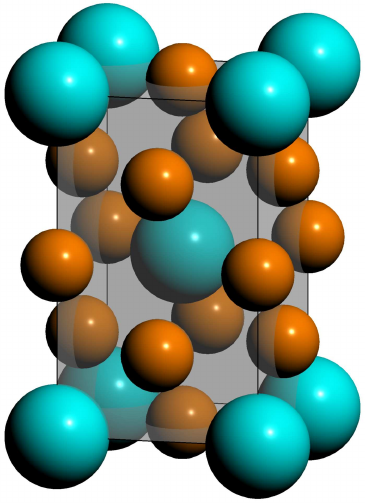}
         \label{fig:D0_22_Cu3W}}
    \qquad
    \subfloat[L1$_0$ CuW]{\includegraphics[width=0.2681\textwidth]{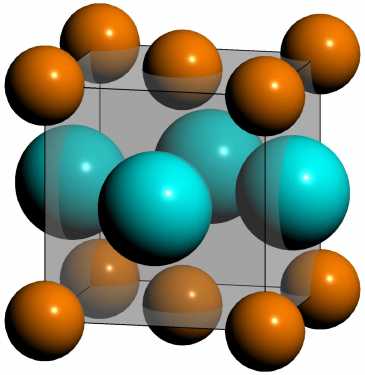}
         \label{fig:L1_0_CuW}}
    \qquad\qquad
    \subfloat[L1$_2$ Cu$_3$W]{\includegraphics[width=0.2100\textwidth]{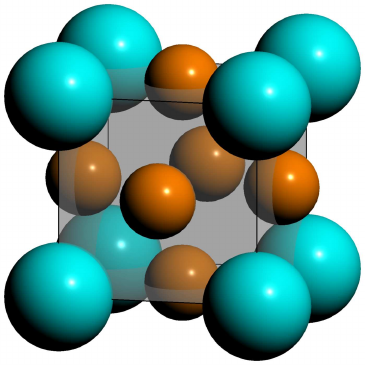}
         \label{fig:L1_2_Cu3W}}
    \quad    
    \caption{Atomic arrangements for possible metastable Cu-W alloys. Cu and W are shown in gold and cyan color spheres, respectively. For (a), (c), (d), and (f), structures with CuW$_3$ stoichiometry are obtained by inverting the Cu and W atoms.}\label{fig:meta_struct}
\end{figure*}

\begin{figure}[htpb]
    \centering
    \subfloat[Cu$_3$W]{\includegraphics[height=0.28\textwidth]{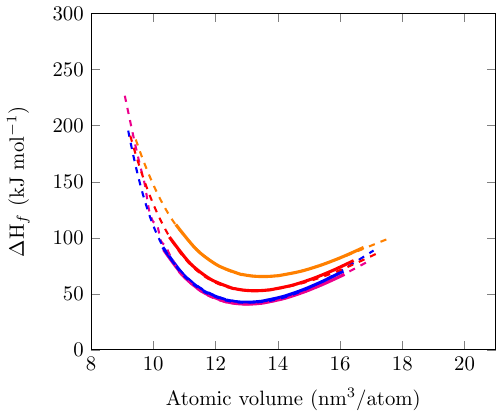}
    \label{fig:metastable_EOS_Cu3W}}
    \subfloat[CuW]{\includegraphics[height=0.28\textwidth]{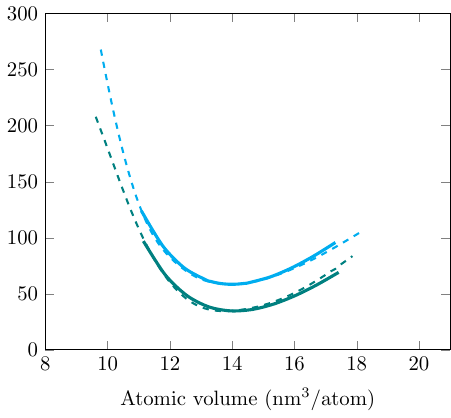}
    \label{fig:metastable_EOS_CuW}}
    \subfloat[CuW$_3$]{\includegraphics[height=0.28\textwidth]{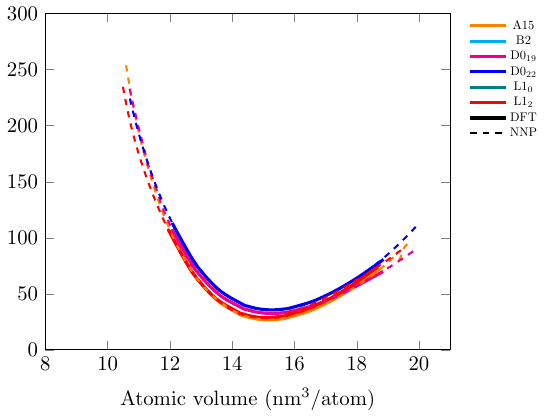}
    \label{fig:metastable_EOS_CuW3}}
    \caption{\label{fig:metastable_EOS}The heats of formation ($\mathrm{\Delta H}_f$)-volume curves for the metastable phases. DFT values are shown with the thick continuous lines, and the NNP values are shown in the dashed lines. The volumes correspond to the volumes of structures in \cref{fig:meta_struct}.}
\end{figure}

Intermetallics of Cu and W are meticulously achieved through non-equilibrium techniques such as ion beam mixing \cite{Zhang2003}.  We have investigated the structures depicted in \cref{fig:meta_struct}, which are candidates for ordered metastable alloys in Cu-W \cite{Zhang2003,Zhang2004,Kong2016}. These structures are included in the reference training dataset.  We have computed the heat of formation ($\mathrm{\Delta H}_f$) versus alloy volume, where $\mathrm{\Delta H}_f$ is calculated as
\begin{equation}\label{eq:e_f}
    \mathrm{\Delta H}_f = \frac{\mathrm{E}_{\mathrm{Cu}_x\mathrm{W}_y}-x\mathrm{E}_\mathrm{Cu}-y\mathrm{E}_\mathrm{W}}{x+y}
\end{equation}
where E$_{\mathrm{Cu}_x\mathrm{W}_y}$ is the energy of a metastable Cu-W solid solution, and E$_\mathrm{Cu}$ and E$_\mathrm{W}$ are the energies of atomic Cu and W in their equilibrium bulk configuration, fcc and bcc for Cu and W, respectively.  The NNP aligns remarkably well with DFT, as shown in \cref{fig:metastable_EOS}. For Cu$_3$W, D0$_{19}$ is the most stable structure closely followed by D0$_{22}$. For CuW, fct L1$_2$ shows significant stability relative to bcc-like B2. For CuW$3$, $\mathrm{\Delta H}_f$ is lower than for other compositions and all structures are much closer in formation energy with A15 being lowest. This provides confidence in applications of the NNP to experimentally-observed interface diffusion and intermixing at the interfaces \cite{Melmed1965,Taylor1966,Wen2007,Zeng2021}. 

\begin{figure}[htpb]
    \centering
    \includegraphics[width=0.7\textwidth]{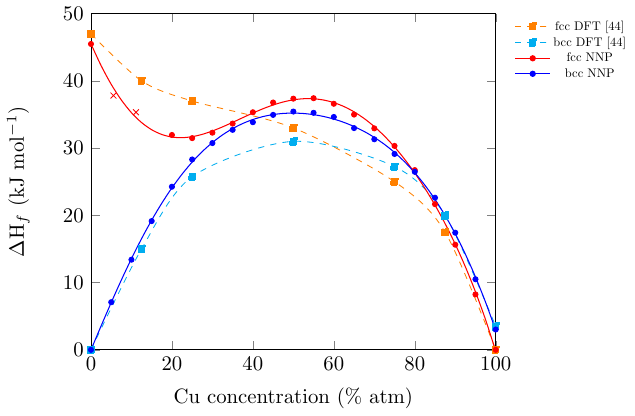}
    \caption{\label{fig:hof_solid_solution}$\mathrm{\Delta H}_f$ for disordered Cu$_x$W$_{100-x}$ solid solutions obtained with the NNP. Data points for $10 \times 10 \times 10$ and $3 \times 3 \times 3$ simulation sizes are  shown in * and $\times$ respectively. The best-fit lines for these data points are also included. $\mathrm{\Delta H}_f$ for DFT calculations from \citet{Liang2017} are also shown here with values for structures obtained with the special quasirandom structure (SQS) method are shown * and cluster expansion values are shown in dashed lines.}
\end{figure}

Metastable solid solutions of Cu-W are mainly observed in Cu-rich fcc structures and W-rich bcc structures. Experimentally, the bcc$\rightarrow$fcc transformation is observed for $>70$\% Cu. We thus examined random Cu$_x$W$_{100-x}$ solid solutions in both fcc and bcc structures.  Simulation cell sizes of $10\times10\times10$ unit cells were relaxed isotropically to preserve the initial crystal structure.  For $x<20$\% Cu, the system was unstable in the fcc phase and so the simulation size was reduced to $3\times3\times3$ unit cells to retain the fcc phase.  The computed $\mathrm{\Delta H}_f$ are shown in \cref{fig:hof_solid_solution}.  The fcc phase becomes energetically stable for $>80$\% Cu, in reasonable agreement with experiments.  However, the NNP shows some differences versus the DFT-computed $\mathrm{\Delta H}_f$  by \citet{Liang2017} using single realizations of 16-atom special quasirandom structure (SQS) (see \cref{fig:hof_solid_solution}). DFT shows the bcc/fcc transition $60$\% Cu, also differing slightly from experiments.  More broadly, the bcc-fcc energy difference versus composition differs between NNP and DFT, especially in the range $20-70$\% Cu, but the  NNP shows good agreement for the formation energies of bcc Cu$_2$W$_{14}$ (12.5\% Cu atm) and Cu$_{14}$W$_2$ (87.5\% Cu atm) \citet{Liang2017}.  Some differences between NNP and DFT can be due to the small size of the DFT simulation cells. 

Overall, we conclude that the NNP should be very useful in studying many aspects of metastable Cu-W alloy behavior and phenomena.

\subsubsection{Cu-W interface structures}

Cu-W interfaces are the most critical features in binary Cu and W since they exist primarily as immiscible composites under equilibrium conditions. The higher density of interfaces in NML systems can significantly affect the properties of the composites; for example, Cu-W NML composites have been observed to degrade when annealed to temperatures $>400\ ^\circ$C. Further, the experimentally observed interface stresses at Cu-W interfaces show a variation that exceeds expectations of sharp interfaces \cite{Lorenzin2024}. Understanding the underlying atomistic mechanisms is critical in realizing possible applications of these NMLs. Highly incoherent interfaces and nanometric layer thicknesses beyond 2 nm make first-principles simulations impractical \cite{Lorenzin2024}, and hence \textit{Ab initio} calculations in the literature were carried out using very modest-size systems with low complexity.

The interface structures in the reference dataset were fully relaxed (cell shape; ionic coordinates) with DFT and the NNP.  Six different interfaces were considered, with the notation ``Interface $x$” with $x=1,2,3,4,5,6$,  for the interface planes and in-plane dimensions of the simulation cells shown in \cref{tab:interface_definition}. Interfaces 4 and 5 are more critical due to the $(111)_\mathrm{Cu}//(110)_\mathrm{W}$ texture observed in the experiments \cite{Taylor1966,Moss1969,Gadyszewski1994,Cancellieri2016,Moszner2016,Druzhinin2019,Yang2021,Yeom2023} with Interface 4 in the NW orientation and Interface 5 rotated 90$^\circ$ from the NW orientation.  The KS orientation is between interfaces 4 and 5, with an inclination of 5.26$^\circ$ from interface 4 (NW orientation). All interface studies use 8 atomic planes each for layers with $\{100\}$  and $\{110\}$ surfaces and 12 atomic planes for layers with $\{111\}$ surfaces.  The misfit strains of Cu and W ($\varepsilon_\mathrm{Cu}$ and $\varepsilon_\mathrm{W}$ for the in-plane x,y directions in \cref{tab:interface_definition}, with tensile strains taken as positive) and the interface energy (E$_\mathrm{IE}$) were compared between DFT and the NNP (see \cref{tab:properties_interface}). E$_\mathrm{IE}$ was determined as
\begin{equation}
    \mathrm{E_{IE}} = \frac{\mathrm{E_{Cu-W}}-\mathrm{E'_{Cu}}-\mathrm{E'_{W}}}{\mathrm{2A}}
\end{equation}
where $\mathrm{E_{Cu-W}}$ is the energy of the simulation cell structure containing the interface and $\mathrm{E'_{Cu}}$ and $\mathrm{E'_{W}}$ are the energies of bulk Cu and W strained to the dimensions of the relaxed cell containing two interfaces each with area A. 

\begin{table}[htpb]
\centering
\caption{\label{tab:interface_definition}Interface planes and the in-plane dimensions of the simulation cells for the interface structures used in training the NNP. The planes and the dimensions use the Miller index notations of the constituent fcc Cu and bcc W layers.}
    \begin{tabular}{c c c c}
    \hline\hline
    \multirow{2}{*}{Interface}&\multirow{2}{*}{Interface plane - z}&\multicolumn{2}{c}{In-plane dimensions}\\ \cline{3-4}
    && x & y \\
    \hline
Interface 1&$(100)_\mathrm{Cu}//(100)_\mathrm{W}$&$\langle 010 \rangle _\mathrm{Cu} // \langle 010 \rangle _\mathrm{W}$&$\langle 001 \rangle _\mathrm{Cu} // \langle 001 \rangle _\mathrm{W}$\\
Interface 2 &$(100)_\mathrm{Cu}//(100)_\mathrm{W}$	&$\langle 010 \rangle _\mathrm{Cu} // \langle 011 \rangle _\mathrm{W}$&$\langle 001 \rangle _\mathrm{Cu} // \langle 0\bar{1}1 \rangle _\mathrm{W}$\\
Interface 3 &$(110)_\mathrm{Cu}//(110)_\mathrm{W}$&$\langle 001 \rangle _\mathrm{Cu} // \langle 001 \rangle _\mathrm{W}$&$\langle 1\bar{1}0 \rangle _\mathrm{Cu} // \langle 1\bar{1}0 \rangle _\mathrm{W}$\\
Interface 4 &$(111)_\mathrm{Cu}//(110)_\mathrm{W}$	&0.5$\langle \bar{1}10 \rangle _\mathrm{Cu} // \langle 001 \rangle _\mathrm{W}$&0.5$\langle 11\bar{2} \rangle _\mathrm{Cu} // \langle 1\bar{1}0 \rangle _\mathrm{W}$\\
Interface 5 &$(111)_\mathrm{Cu}//(110)_\mathrm{W}$	&$\langle 11\bar{2} \rangle _\mathrm{Cu} // \langle 001 \rangle _\mathrm{W}$&$\langle \bar{1}10 \rangle _\mathrm{Cu} // \langle 1\bar{1}0 \rangle _\mathrm{W}$\\
Interface 6 &$(111)_\mathrm{Cu}//(111)_\mathrm{W}$	&0.5$\langle 11\bar{2} \rangle _\mathrm{Cu} // \langle 1\bar{1}0 \rangle _\mathrm{W}$&$1.5\langle 1\bar{1}0 \rangle _\mathrm{Cu} // \langle 11\bar{2} \rangle _\mathrm{W}$\\
\hline
    \end{tabular}
\end{table}
\begin{table}[htpb]
\centering
\caption{\label{tab:properties_interface}Comparison of the interface energies and the misfit strains obtained for the NNP with the corresponding properties obtained from DFT calculations}
    \begin{tabular}{c c c | c c | c c | c c | c c}
    \hline\hline
    \multirow{2}{*}{Interface}&\multicolumn{2}{c}{E$_\mathrm{IE}$ (mj/m$^{-2}$)}&\multicolumn{2}{c}{$\varepsilon_\mathrm{Cu,x}$ (\%)}&\multicolumn{2}{c}{$\varepsilon_\mathrm{Cu,y}$ (\%)}&\multicolumn{2}{c}{$\varepsilon_\mathrm{W,x}$ (\%)}&\multicolumn{2}{c}{$\varepsilon_\mathrm{W,y}$ (\%)}\\ \cline{2-11}
    &DFT&NNP&DFT&NNP&DFT&NNP&DFT&NNP&DFT&NNP\\
    \hline
Interface 1	&1979	&1973&-6.8&-6.2&-6.8&-6.2&6.1&6.7&6.1&6.7\\
Interface 2	&1189	&1258&22.6&22.2&22.6&22.2&-1.3&-1.7&-1.3&-1.7\\
Interface 3	&1637	&1474&-14.6&-15.4&-7.6&-6.9&-2.8&-3.7&5.1&5.9\\
Interface 4	&844	&830&19.4&19.8&-0.6&-0.5&-3.9&-3.6&-2.0&-1.9\\
Interface 5	&2255	&2189&-5.0&-4.8&-1.2&0.3&8.1&8.3&37.7&39.7\\
Interface 6	&2325	&2296&0.3&0.6&0.3&0.6&-1.2&-0.8&-1.2&-0.8\\
\hline
    \end{tabular}
\end{table}

\begin{figure}[htpb]
    \centering
    \subfloat[Interface 4 - DFT]{\includegraphics[width=0.45\textwidth]{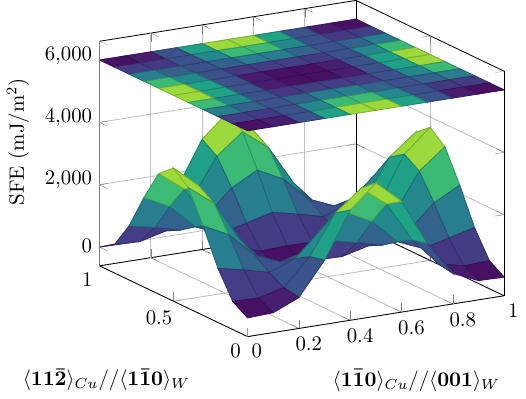}
         \label{fig:DFT_slip_4}}
    \quad
    \subfloat[Interface 4 - NNP]{\includegraphics[width=0.45\textwidth]{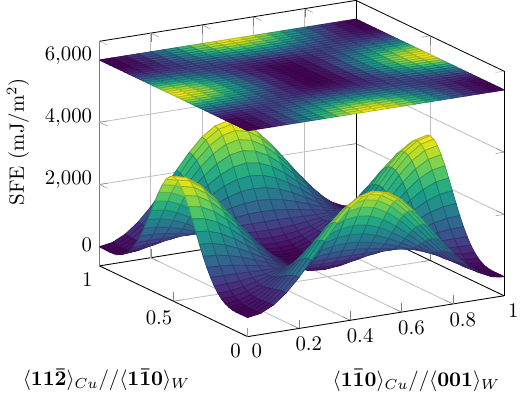}
         \label{fig:NNP_slip_4}}
    \quad
    \subfloat[Interface 5 - DFT]{\includegraphics[width=0.45\textwidth]{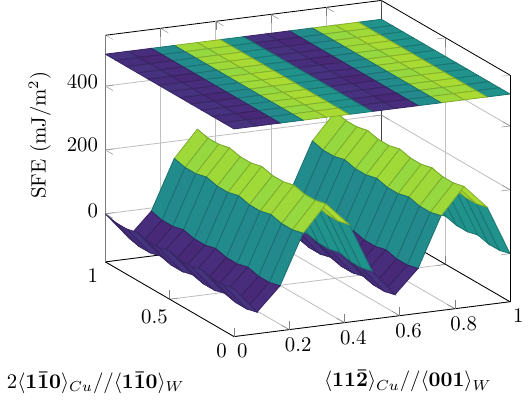}
         \label{fig:DFT_slip_5}}
    \quad
    \subfloat[Interface 5 - NNP]{\includegraphics[width=0.45\textwidth]{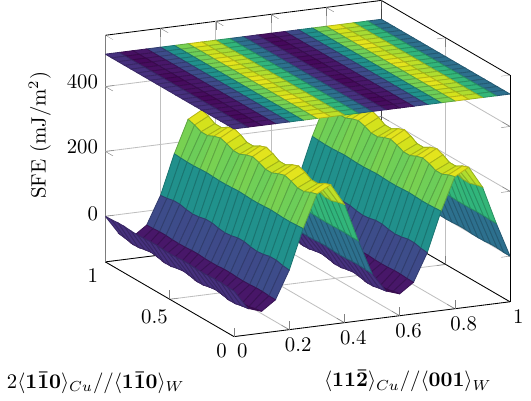}
         \label{fig:NNP_slip_5}}
    \caption{Slip energy surfaces for interfaces 4 and 5 obtained with DFT and the NNP. The energies were obtained for rigid slips at the interface. }\label{fig:slip}
\end{figure}

The largest percentage error is for the interface energies for interfaces 2 and 3, showing 5\% -10\%). However, the agreement for the more critical Interfaces 4 and 5 is excellent.  There is also very good agreement between DFT and NNP for the misfit strains in all the interfaces.  Due to the modest sizes used with DFT, the misfit strains are extremely large.  The only reasonable misfit strains are for Interface 6, which has not been observed in Cu-W systems.  Due to the high misfit strains in these interfaces, real multilayer structures will be incoherent.  Information on local atomic environments in these environments was captured in the reference dataset by providing rigid relative displacements at the interfaces. We compute the slip energy surfaces for interfaces 4 and 5, which captures the behavior of the common $\{111\}_\mathrm{Cu}//\{110\}_\mathrm{W}$ interfaces, with the NNP and DFT and find excellent agreement, as shown in \cref{fig:slip}. Furthermore, the NNP energy surfaces are smooth and continuous. 

\begin{figure}[htpb]
    \centering
    \subfloat[KS orientation for the cells size in \citet{Bodlos2022}]{\includegraphics[width=0.45\textwidth]{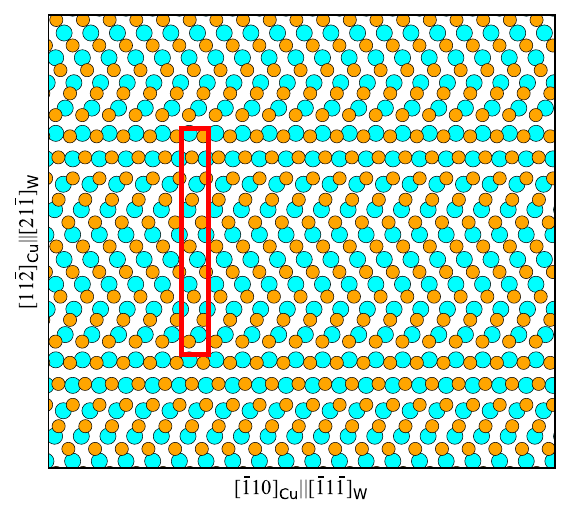}
         \label{fig:KS_bodlos}}
    \quad
    \subfloat[NW orientation]{\includegraphics[width=0.45\textwidth]{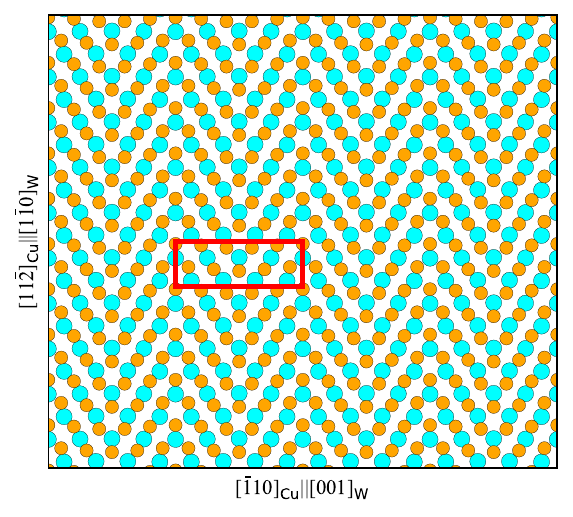}
         \label{fig:NW_bodlos}}
    \quad
    \subfloat[KS orientation for the cells size with lower misfit strain]{\includegraphics[width=0.8\textwidth]{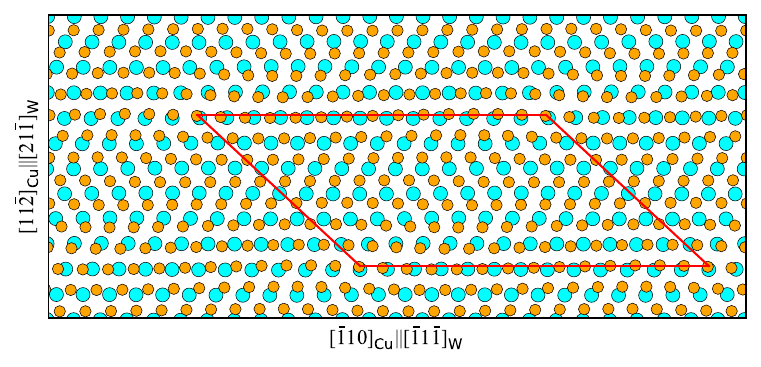}
    \label{fig:KS_NNP01}}
     \caption{Atomic arrangements of the interfaces obtained with the NNP. Only the two atomic layers adjacent to the interfaces are shown for clarity, with the repeating unit cell identified by with the red square.}\label{fig:Boldos_interfaces}
\end{figure}

The interface structures in the reference dataset have high misfit strains (the largest being 37\%-40\% strain in the W layer for interface 5) due to the modest size of the simulation cells used to compensate for the computational costs of DFT. However, realistic interfaces will have much smaller misfit strains to reduce the strain energy of the Cu and W layers, which can have thicknesses in the nm range. To further validate the ability of the NNP to model $\{111\}_\mathrm{Cu}//\{110\}_\mathrm{W}$ interfaces, we compare with results from \citet{Bodlos2022}, who simulated $\{111\}_\mathrm{Cu}//\{110\}_\mathrm{W}$ interfaces with DFT for larger in-plane dimensions for the NW and KS orientations, albeit with a single Cu layer and a single W layer connected through a single interface. Each layer was six atomic planes thick, while the surfaces opposite the interface faced vacuums. Using the NNP, the misfit strains, interface energies (E$_\mathrm{IE}$), and work of separation (E$_\mathrm{WoS}$), and the results are shown in \cref{tab:interface_prop} along with the DFT results of \citet{Bodlos2022} (not used in training). The E$_\mathrm{IE}$ and E$_\mathrm{WoS}$ are in good agreement, and some errors could be due to differences in the DFT parameters \citet{Bodlos2022} versus those used here for training the NNP.  The atomic arrangements adjacent to the interfaces for these structures as computed by the NNP are shown in \cref{fig:KS_bodlos} and \cref{fig:NW_bodlos} for the KS and NW orientations, respectively. These arrangements match the DFT results in \citet{Bodlos2022}. These results further confirm the ability of the NNP to model the interfaces with close to DFT accuracy. 

\begin{table}[htpb]
\centering
\caption{\label{tab:interface_prop}E$_\mathrm{IE}$, E$_\mathrm{WoS}$, $\varepsilon_\mathrm{Cu}$, and $\varepsilon_\mathrm{W}$,  obtained for KS and NW interfaces with NNP. DFT values from \citet{Bodlos2022} are shown when available.}
    \begin{tabular}{l c c c| c c c}
    \hline\hline
    \multirow{2}{*}{Property}&\multicolumn{3}{c}{KS interface}&\multicolumn{3}{c}{NW interface}\\ \cline{2-7}
    & DFT \cite{Bodlos2022} & NNP\footnotemark[1] & NNP\footnotemark[2] & DFT \cite{Bodlos2022}& NNP\footnotemark[1] & NNP\footnotemark[2] \\
    \hline
E$_\mathrm{IE}$ (mJ/m$^2$) & 1320 & 1278 & 1392 & 1170 & 1254 & 1289\\
E$_\mathrm{WoS}$ (mJ/m$^2$) &3220&3214&-&3400&3275&-\\
$\varepsilon_\mathrm{Cu} (x, y)$ (\%)&-&5.0,2.9&-0.2,0.2&-&0.6,-2.1&-0.8,1.2\\
$\varepsilon_\mathrm{W}  (x, y)$ (\%) &-&-2.4,-2.4&-0.1,-0.2&-&-0.8,-1.5&-0.1,-0.3\\
\hline
\footnotetext[1]{For the simulation cell size in \citet{Bodlos2022}}
\footnotetext[2]{For the simulation cell size with lower misfit strain}
   \end{tabular}
\end{table}

The misfit strains of these structures (see \cref{tab:interface_prop}) could still be quite large compared to realistic systems. This can cause considerable strain energies, especially since the layer thicknesses of realistic NMLs are large (in the nm range). Therefore, we simulated more realistic NMLs for NW and KS interfaces with alternating Cu and W layers of $\sim5$ nm thicknesses. The misfit strains for these structures are reduced compared to the structures from \citet{Bodlos2022}. For the NW interface, where only the layer thicknesses and periodicity normal to the interfaces are changed, there is a marginal increase in the interface energy (2.8\%) and no significant change in structure compared to the difference from \cref{fig:NW_bodlos}.  However, the E$_\mathrm{IE}$ for the KS orientation is noticeably increased (9\%) and the atomic arrangement adjacent to the KS interface in the larger simulation cell is more complex (see \cref{fig:KS_NNP01}) as compared to \cref{fig:KS_bodlos}.  These changes in interface structure and misfit strains in the layers due to the use of realistic layer thickness demonstrate that understanding complex interface behavior in real systems, including microstructure evolution (from NML to nanocomposite), intermixing, and delamination, requires such large sizes, which in turn requires the use of the NNP.

 With stability, near-DFT accuracy, and reasonable computational costs, this Cu/W NNP is poised to enable the identification of the underlying atomistic origins of interface structures, energies, and stresses.  Progress in this direction is discussed in forthcoming work \cite{Hu2024}.

\subsubsection{Interactions between solid W and liquid Cu}

\begin{figure}[htpb]
    \centering
    \subfloat[Liquid Cu sandwiched between $\{110\}$ W surfaces]{\includegraphics[width=0.365\textwidth]{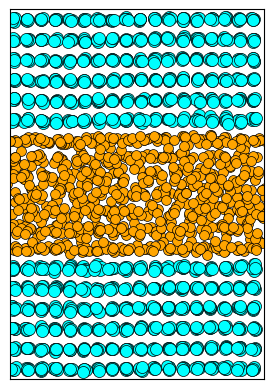}
         \label{fig:solid_liquid_1}}
    \quad
    \subfloat[W precipitate suspended in liquid Cu]{\includegraphics[width=0.516\textwidth]{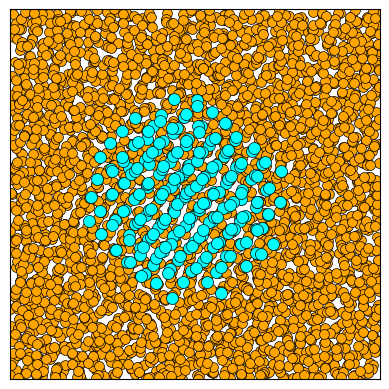}
         \label{fig:solid_liquid_2}}
     \caption{Typical atomic arrangements observed for liquid Cu and solid W systems under equilibrium. Gold and cyan circles represent Cu and W, respectively.}\label{fig:solid_liquid}
\end{figure}

Due to a positive enthalpy of mixing, the formation of Cu and W solutions is discouraged. Differences in the melting points of Cu and W facilitate the infiltration of liquid Cu into porous W matrices when manufacturing Cu-W MMCs. Hence, we investigate the ability of the NNP to simulate solid W/liquid Cu interfaces.  We study four configurations: (i) a liquid Cu layer ($\sim15$ \r{A} thick) sandwiched between bcc W layers with $\{110\}$ surfaces at 1600 K; (ii) a spherical W precipitate (radius of 20 \r{A}) suspended in liquid Cu at 1600 K; (iii) separation of bcc W in an initial liquid Cu-W mixture (95\% Cu and 5\% W) at 1600 K; and (iv) wetting angles for Cu in a W (110) surface at 1350 K. Simulations for configurations (i) and (ii) use the NPT ensemble with 0.0005 ps timestep for 150 ps with Cu and W in liquid and solid states, respectively. \Cref{fig:solid_liquid} shows snapshots of typical arrangements after the systems reach equilibrium. We see no changes in the Cu or W phases during the simulations. The NNP simulates these structures well, with small deviations around the mean and no irregular energy variations.

\begin{figure}[htpb]
    \centering
    \subfloat[t = 0]{\includegraphics[width=0.22\textwidth]{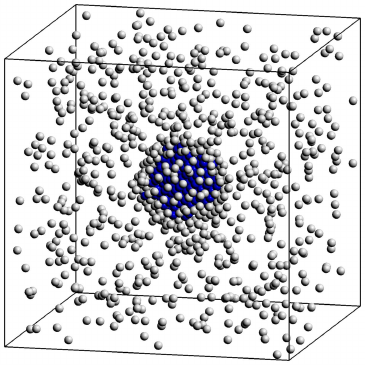}
         \label{fig:t_0}}
    \quad
    \subfloat[t = 400 ps]{\includegraphics[width=0.22\textwidth]{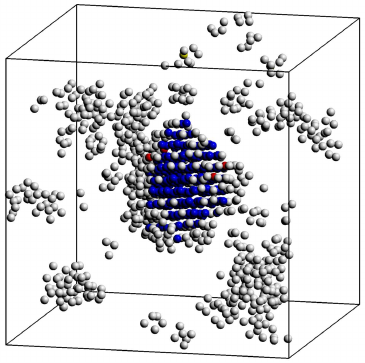}
         \label{fig:t_400}}
    \quad
    \subfloat[t = 4000 ps]{\includegraphics[width=0.22\textwidth]{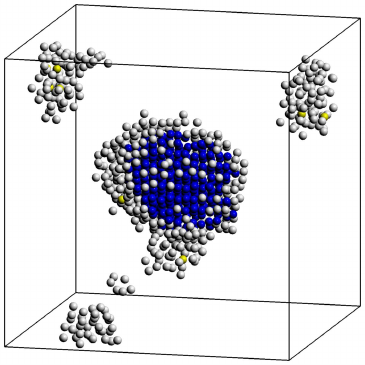}
         \label{fig:t_4000}}
    \quad
    \subfloat[t = 8000 ps]{\includegraphics[width=0.22\textwidth]{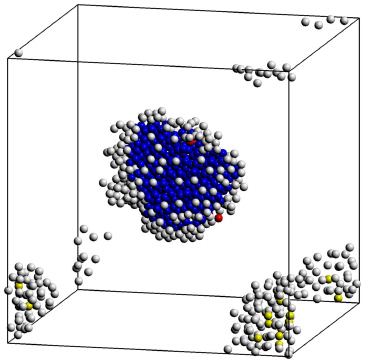}
         \label{fig:t_8000}}
     \caption{Precipitation of W in a liquid mixture of Cu and W. Only the W atoms are shown here. Blue, red,  and yellow circles represent atoms in bcc, hcp, and icosahedral crystal structures, respectively, with white circles representing atoms that cannot be classified to any structural system as identified by common neighbor analysis \cite{Honeycutt1987,Faken1994}}\label{fig:precipitation}
\end{figure}

The separation of bcc W from the Cu-W liquid mixture is initiated by introducing a small spherical bcc W precipitate (radius of 10 \r{A}) to a liquid Cu-W mixture (5\% W). The simulation is executed in the NPT ensemble with 0.002 ps timestep at 1600 K for a period of 8000 ps. \Cref{fig:precipitation} shows snapshots of the system at different times. Initially, the W atoms aggregate to form smaller precipitates. As the system evolves, the smaller precipitates aggregate to form larger aggregates while the initial bcc W precipitate grows.  The crystal structures of the small precipitates cannot be classified, which may be due to artifacts in the potential or simply the large number of interface atoms. However, the bcc precipitate grows steadily while the system energy is decreasing.  The surfaces of the final bcc W precipitate are dominated by $\{110\}$ planes, as expected, since this is the lowest-energy surface of bcc W.

\begin{figure}[htpb]
    \centering
    \subfloat[Normal to $\langle001\rangle$ in the W surface]{\includegraphics[width=0.8\textwidth]{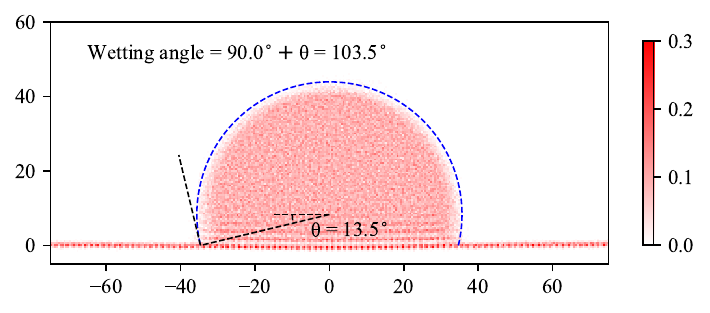}
         \label{fig:wetting_001}}
    \quad
    \subfloat[Normal to $\langle\bar{1}10\rangle$ in the W surface]{\includegraphics[width=0.8\textwidth]{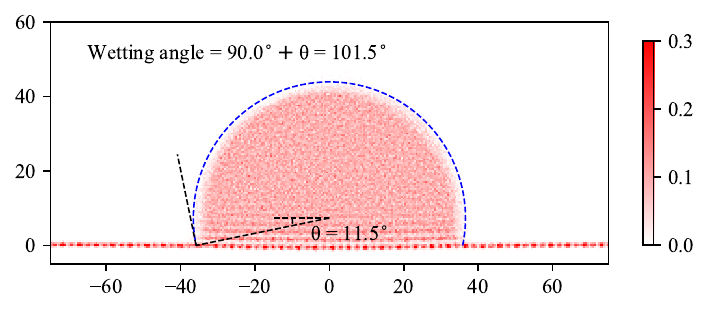}
         \label{fig:wetting_-110}}
    \caption{\label{fig:wetting_angle}The probability distribution of Cu atoms for a 5 \r{A} thick layer at the center of mass of the Cu droplet. The space was meshed with $0.5\times0.5$ \r{A}$^2$ units. A circle was fitted to the surface of the Cu droplet with the blue color dashed curve.}
\end{figure}

Finally, we simulate the wetting angle of a Cu droplet on a (110) W surface. A spherical Cu droplet with 8000 atoms is placed on a $\sim15$ \r{A} thick W layer with $\sim190\times 190$ \r{A}$^2$ surface area. The simulation is executed in the NPT ensemble with 0.002 ps timestep at 1350 K for 350 ps.  The system reaches a quasi-equilibrium state in 40 ps. \Cref{fig:wetting_angle} shows the probability distribution of Cu atoms for 5 \r{A} thick layers in the orthogonal $\langle001\rangle$ and $\langle\bar{1}10\rangle$ directions in the (110) W surface. The wetting angles for the two directions are very similar (103.5$^\circ$; 101.5$^\circ$) but higher than the observed stabilized wetting angle \cite{Izaguirre2024}.  The system also forms a one-atom-thick ``precursor” Cu film spread over the entire W (110) surface, for which there is no evidence in Cu-W.

The Cu monolayer on (110) W may be driven by a Cu adsorption energy that is stronger in the NNP then in DFT.  The adsorption energy of a single Cu atom on the (110) surface (E$_\mathrm{ads}$) is calculated at 0 K as
\begin{equation}
    \mathrm{E}_\mathrm{ads} = \mathrm{E}_\mathrm{sys} - \mathrm{E}_\mathrm{slab} - \mathrm{E}_\mathrm{Cu}
\end{equation}
where E$_\mathrm{sys}$, E$_\mathrm{slab}$, and E$_\mathrm{Cu}$ are the energies of the system with an adsorbed Cu atom, the system with clean surfaces, and a Cu atom in fcc bulk system. E$_\mathrm{ads}$ computed from DFT and NNP are -0.033 eV and -0.378 eV, respectively. The much lower NNP value drives the formation of the precursor Cu monolyaer film and then likely influences the observed wetting angles.  The energetics of adsorbed Cu on W surfaces are not represented in the reference dataset for the current NNP.  If Cu wetting becomes an important application issue, then a new NNP can be created that incorporates the relevant DFT information in the training data.   

\section{Conclusions\label{sec:conclusions}}

Composites of Cu and W have recently gained significant attention and have potential applications in brazing fillers and shielding materials for plasma and radiation. The immiscibility of Cu-W composites leads to their fabrication as nano-multilayers, nano-composites, and metal matrix composites, sparking interest in understanding the behavior of interfaces in the solid state and between liquid W and solid Cu. We have developed a neural network potential using the Behler-Parrinello framework to enable the detailed study of atomistic mechanisms associated with these Cu-W materials. This potential, created with a meticulously curated reference dataset and an optimized set of descriptors, provides very good accuracy versus first-principles methods while overcoming their computational limitations.

The NNP was rigorously validated for the mechanical and thermodynamic behavior of bulk Cu and W. We also assessed its performance for binary Cu-W systems, including substitutions, metastable alloys, and standard Cu-W interfaces. The thermodynamics of binary systems, particularly the interactions between solid W and liquid Cu, were also examined. The NNP demonstrates excellent quality in these studies, showing no instabilities and maintaining good agreement with first-principles values and experimental observations where applicable. This successful validation instills confidence in the reliability and accuracy of the NNP. 

The ability of the NNP to predict a comprehensive range of behavior opens up many new possibilities for modeling interfaces.  In particular, the NNP can be used to investigate intermixing, phase stability, and the large observed variation in interface stresses versus processing conditions, enhancing our understanding of these phenomena; we will report on progress in this direction in forthcoming work \cite{Hu2024}. The Cu-W NNP can also elucidate the effect of layer thickness, in-plane stress, macroscopic porosity, temperature, grain boundaries, and non-equilibrium vacancies in physical vapor-deposited Cu/W nano-multilayers on Young's modulus, explaining contradictory experimental measurements in relation to different deposition conditions and resulting microstructures; we will report on progress in this direction in another forthcoming work \cite{Lorenzin-elastic2024}.  More broadly, the versatility of this NNP, and others that have been developed recently, demonstrates that machine-learned potentials provide a valuable tool for studying many atomic-scale problems in materials science and engineering.

\begin{acknowledgments}
This research was supported by the NCCR MARVEL, a National Centre of Competence in Research, funded by the Swiss National Science Foundation (grant number 205602). 
\end{acknowledgments}

\appendix
\section{\label{app:sym_hyp}Symmetry function hyper-parameters used in developing Cu-W NNP potential}

Hyper-parameters of the symmetry functions used when developing the Cu-W potential are shown in \cref{tab:rad_syms,tab:ang_syms}. Symmetry function types 2, 3, and 9 correspond to $G^2$, $G^3$, and $G^9$ symmetry functions as shown in \cref{eq:radial,eq:narrow,eq:wide} respectively.

\begin{longtable}[c]{c c c c}
    \caption{\label{tab:rad_syms}Hyper-parameters for the radial symmetry functions ($G^2$) selected by the PCov-CUR decomposition algorithm \cite{Cersonsk2021} as described in \cref{eq:radial}. $r_\textrm{c}=7.0$ for all the symmetry functions}\\
    \hline\hline
    Central &  Neighbor & \multirow{2}{*}{$\eta$ (\r{A}$^{-2}$)} & \multirow{2}{*}{$r_\textrm{s}$ (\r{A})}\\
    atom type & atom type & & \\
    \hline
    \endfirsthead
    \caption{(Continued.)}\\
    \hline\hline
    Central &  Neighbor & \multirow{2}{*}{$\eta$ (\r{A}$^{-2}$)} & \multirow{2}{*}{$r_\textrm{s}$ (\r{A})}\\
    atom type & atom type & & \\    \hline
    \endhead
    \hline
    \endfoot
    \hline
    \endlastfoot
Cu&W&2.0000&1.50\\
Cu&W&2.0000&2.00\\
Cu&W&2.0000&2.50\\
Cu&W&2.0000&3.00\\
Cu&W&2.0000&3.50\\
Cu&W&2.0000&4.00\\
Cu&W&2.0000&4.50\\
Cu&W&2.0000&5.00\\
Cu&W&2.0000&5.50\\
Cu&W&2.0000&6.00\\
Cu&Cu&2.0000&2.00\\
Cu&Cu&2.0000&2.50\\
Cu&Cu&2.0000&3.00\\
Cu&Cu&2.0000&3.50\\
Cu&Cu&2.0000&4.00\\
Cu&Cu&2.0000&4.50\\
Cu&Cu&2.0000&5.00\\
Cu&Cu&2.0000&5.50\\
Cu&Cu&2.0000&6.00\\
Cu&Cu&2.0000&6.50\\
Cu&Cu&0.0204&0.00\\
W&W&2.0000&2.00\\
W&W&2.0000&2.50\\
W&W&2.0000&3.00\\
W&W&2.0000&3.50\\
W&W&2.0000&4.00\\
W&W&2.0000&4.50\\
W&W&2.0000&5.00\\
W&W&2.0000&5.50\\
W&W&2.0000&6.00\\
W&W&2.0000&6.50\\
W&W&0.0204&0.00\\
W&Cu&2.0000&2.00\\
W&Cu&2.0000&2.50\\
W&Cu&2.0000&3.00\\
W&Cu&2.0000&3.50\\
W&Cu&2.0000&4.00\\
W&Cu&2.0000&4.50\\
W&Cu&2.0000&5.00\\
W&Cu&2.0000&5.50\\
W&Cu&2.0000&6.00\\
W&Cu&0.0204&0.00\\

\end{longtable}

\begin{longtable}{c c c c c c c}
    \caption{\label{tab:ang_syms}Hyper-parameters for the angular symmetry functions ($G^3$ and $G^9$) for selected by the PCov-CUR decomposition algorithm \cite{Cersonsk2021} as described in \cref{eq:narrow,eq:wide}. $r_\textrm{c}=7.0$ for all the symmetry functions}\\
    \hline\hline
    Central & \multirow{2}{*}{Type} & Neighbor & Neighbor & \multirow{2}{*}{$\eta$ (\r{A}$^{-2}$)} & \multirow{2}{*}{$\lambda$} & \multirow{2}{*}{$\zeta$} \\
    atom type & & atom 1 type & atom 2 type & & & \\
    \hline
    \endfirsthead
    \caption{(Continued.)}\\
    \hline\hline
    Central & \multirow{2}{*}{Type} & Neighbor & Neighbor & \multirow{2}{*}{$\eta$ (\r{A}$^{-2}$)} & \multirow{2}{*}{$\lambda$} & \multirow{2}{*}{$\zeta$} \\
    atom type & & atom 1 type & atom 2 type & & & \\
    \hline
    \endhead
    \hline
    \endfoot
    \hline
    \endlastfoot
Cu&9&W&W&0.0204&1&1\\
Cu&9&W&W&0.0204&1&3\\
Cu&9&W&W&0.0204&-1&1\\
Cu&9&W&W&0.0204&-1&12\\
Cu&9&W&W&0.0204&-1&64\\
Cu&9&W&W&0.0740&1&1\\
Cu&9&Cu&W&0.1408&-1&1\\
Cu&9&Cu&W&0.0204&1&1\\
Cu&9&Cu&W&0.0204&1&3\\
Cu&9&Cu&W&0.0204&1&12\\
Cu&9&Cu&W&0.0204&1&64\\
Cu&9&Cu&W&0.0204&-1&1\\
Cu&9&Cu&W&0.0204&-1&12\\
Cu&9&Cu&W&0.0204&-1&64\\
Cu&9&Cu&W&0.0740&1&1\\
Cu&9&Cu&W&0.0740&-1&1\\
Cu&9&Cu&W&0.0740&-1&12\\
Cu&9&Cu&Cu&0.0204&1&1\\
Cu&9&Cu&Cu&0.0204&1&3\\
Cu&9&Cu&Cu&0.0204&1&12\\
Cu&9&Cu&Cu&0.0204&1&64\\
Cu&9&Cu&Cu&0.0204&-1&1\\
Cu&9&Cu&Cu&0.0204&-1&12\\
Cu&9&Cu&Cu&0.0204&-1&64\\
Cu&9&Cu&Cu&0.0740&1&1\\
Cu&9&Cu&Cu&0.0740&-1&1\\
Cu&9&Cu&Cu&0.0740&-1&64\\
W&9&W&W&0.0204&1&1\\
W&9&W&W&0.0204&1&12\\
W&9&W&W&0.0204&1&64\\
W&9&W&W&0.0204&-1&1\\
W&9&W&W&0.0204&-1&3\\
W&9&W&W&0.0204&-1&12\\
W&9&W&W&0.0204&-1&64\\
W&9&W&W&0.0740&1&1\\
W&9&W&W&0.0740&-1&1\\
W&9&W&W&0.0740&-1&64\\
W&9&Cu&W&0.1408&-1&1\\
W&9&Cu&W&0.0204&1&1\\
W&9&Cu&W&0.0204&1&3\\
W&9&Cu&W&0.0204&1&12\\
W&9&Cu&W&0.0204&1&64\\
W&9&Cu&W&0.0204&-1&1\\
W&9&Cu&W&0.0204&-1&12\\
W&9&Cu&W&0.0204&-1&64\\
W&9&Cu&W&0.0740&1&1\\
W&9&Cu&W&0.0740&-1&1\\
W&9&Cu&Cu&0.0204&1&1\\
W&9&Cu&Cu&0.0204&1&3\\
W&9&Cu&Cu&0.0204&1&12\\
W&9&Cu&Cu&0.0204&-1&1\\
W&9&Cu&Cu&0.0204&-1&12\\
W&9&Cu&Cu&0.0204&-1&64\\
W&9&Cu&Cu&0.0740&-1&1\\
\end{longtable}
\bibliography{references}

\end{document}